\documentclass[aps,prl,10pt,twocolumn,showpacs,preprintnumbers,amsmath,amssymb,superscriptaddress]{revtex4-1}
\usepackage[english]{babel}
\usepackage[dvips]{graphicx}
\usepackage{dcolumn}
\usepackage{bm}

\usepackage{color}

\begin{document}

\author{M. E. Dieckmann}
\affiliation{Department of Science and Technology (ITN), Link\"opings University, Campus Norrk\"oping, SE-60174 Norrk\"oping, Sweden}
\author{A. Alejo}
\affiliation{Centre for Plasma Physics (CPP), Queen's University Belfast, BT7 1NN, UK}
\author{G. Sarri}
\affiliation{Centre for Plasma Physics (CPP), Queen's University Belfast, BT7 1NN, UK}

\date{\today}
\pacs{52.65.Rr,52.72.+v,52.27.Ep}

\email{mark.e.dieckmann@liu.se}
\title{Expansion of a mildly relativistic hot pair cloud into an electron-proton plasma}
\begin{abstract}
The expansion of a charge-neutral cloud of electrons and positrons with the temperature 1 MeV into an unmagnetized ambient plasma is examined with a 2D particle-in-cell (PIC) simulation. The pair outflow drives solitary waves in the ambient protons. Their bipolar electric fields attract electrons of the outflowing pair cloud and repel positrons. These fields can reflect some of the protons thereby accelerating them to almost an MeV. Ion acoustic solitary waves are thus an efficient means to couple energy from the pair cloud to protons. The scattering of the electrons and positrons by the electric field slows down their expansion to a nonrelativistic speed. Only a dilute pair outflow reaches the expansion speed expected from the cloud's thermal speed. Its positrons are more energetic than its electrons. In time an instability grows at the front of the dense slow-moving part of the pair cloud, which magnetizes the plasma. The instability is driven by the interaction of the outflowing positrons with the protons. These results shed light on how magnetic fields are created and ions are accelerated in pair-loaded astrophysical jets and winds.
\end{abstract}

\maketitle

\section{Introduction}

Intense electromagnetic fields and radiation close to compact astrophysical objects, such as neutron stars and black holes, trigger the formation of dense clouds of electrons and positrons \cite{Ruffini10}. It has been proposed that pair clouds could emerge in regions with a low plasma density, which are known as gaps, where electric fields can accelerate electrons to energies that are sufficiently high to trigger pair formation via collisions. Recent global particle-in-cell (PIC) simulations have examined how pairs fill the magnetosphere of the compact object and escape along open field lines \cite{Cerutti17} forming the pulsar wind. 

The radiation associated with the pair cloud ionizes the material along its path. The particle's mean-free path in these dilute plasmas is large and binary collisions with the ambient medium may not be an efficient means to slow down and to thermalize the pair clouds. The clouds interact in this case with the ambient plasma via collisionless plasma instabilities, which lead to the growth of electromagnetic fields. These fields redistribute energy between the plasma particles, which can result in the acceleration of some particles to high energies. It is important to identify the instabilities that grow and the nonlinear plasma structures that form under these extreme conditions and how they accelerate particles \cite{Marcowith16}. 

Laboratory experiments can shed light on how energetic pair clouds interact with an ambient plasma, which consists of electrons and ions. Neutral and dense clouds of electrons and positrons can be created in the laboratory using high-power laser pulses \cite{Chen09,DiPiazza12,Sarri13,Sarri15,Liang15}. The initial temperature and mean speed of the pair cloud is such that the characteristic kinetic energy of the leptons is at least comparable to their rest mass energy in the rest frame of the cloud. The pair cloud propagates through a low density electron-proton plasma resulting from photo-ionisation of the residual gas in the vacuum chamber. The high temperature of the plasma implies that binary collisions between plasma particles are negligible and the plasma behaves as a collisionless medium. The plasma processes observed in such a plasma may thus be similar to those found in its astrophysical counterpart \cite{Warwick18}.

Pair clouds and the kinetic energy they carry are presently small and their interaction with the ambient plasma can not accelerate many ambient ions to large speeds. The pair cloud is however dense enough to interact with the ambient electrons via collective processes. If the pair cloud propagates at a speed relative to the ambient plasma that is large compared to its thermal speed then it drives a beam-Weibel instability \cite{Bret10,Sarri17}. This instability triggers the growth of magnetic fields in the ambient plasma, it heats up the ambient electrons and it scatters the cloud particles. 

Pair clouds close to compact astrophysical objects and, potentially, those generated by forthcoming more powerful lasers should be large enough to accelerate ions. One-dimensional PIC simulations \cite{Dieckmann18} have examined the interaction of a hot pair cloud that expands into an ambient plasma. The simulations have shown that electron phase space holes \cite{Morse69,Schamel86,Luque05,Eliasson05}, which form and evolve on electron time scales, and the ion acoustic solitary waves that are tied to them \cite{Jenab17} could be an efficient means to accelerate ions to high energies especially when the ion acoustic solitary waves break \cite{Liseykina15}. Our aim is to study these processes in more detail with a two-dimensional particle-in-cell (PIC) simulation that resolves competing multidimensional instabilities such as those that destroy phase space holes in more than one dimension \cite{Morse69}. 

The pair cloud in our simulation is initially at rest and its thermal pressure lets it expand into the ambient medium, which consists of electrons and protons. We observe the growth of proton velocity oscillations, which we interpret as ion acoustic solitary waves. Protons are accelerated to even higher energies at the crests of some of these waves. The observed bipolar electric field pulses suggest the involvement of electron phase space holes in the proton acceleration to MeV energies also in the two-dimensional simulation. Indeed an ion acoustic solitary wave is tied to an electron phase space hole and stabilizes it \cite{Jenab17}. A second instability emerges at late times. Our simulation results suggest that it is a filamentation instability between the protons and the positrons. It results in a localized evacuation of protons and in the generation of a magnetic field. We list the initial conditions for the simulation in Section 2. Section 3 presents the simulation results, which are summarized in section 4.


\section{Simulation setup}

We employ the PIC simulation code EPOCH \cite{Arber15}. It solves Faraday's and Amp\`ere's law for the electromagnetic fields with a charge-conserving scheme and it approximates the plasma by an ensemble of computational particles (CPs). We do not represent binary collisions in our simulation and the plasma dynamics is thus determined exclusively by the collective electromagnetic fields. 

Our simulation setup is the following. An ambient plasma, which consists of electrons and protons with the density $n_0$, fills the simulation box uniformly. The ambient electrons have the temperature $T_0=$ 1 keV and the protons the temperature $T_0/10$. Such temperatures are typical for the ionized residual gas in laser-plasma experiments on time scales below 1 ns if an ultra-intense laser pulse was employed \cite{Ahmed13}. We normalize space to the proton skin depth $\lambda_s = c/\omega_{pi}$ where $c$ is the light speed and $\omega_{pi}={(n_0e^2/m_p\epsilon_0)}^{1/2}$ the plasma frequency of the ambient protons ($e, m_p, \epsilon_0:$ elementary charge, proton mass and vacuum permittivity). Time is normalized to $\omega_{pi}^{-1}$. The electron plasma frequency is $\omega_{pe}={(m_p/m_e)}^{1/2}\omega_{pi}$ with $m_p/ m_e =1836$. The thermal speeds of the ambient electrons and protons are $v_{e,th}={(k_BT_0/m_e)}^{1/2} \approx 1.3 \times 10^7$ m/s and $v_{p,th}={(k_BT_0/10 m_p)}^{1/2}\approx 10^5$ m/s, respectively. The ambient plasma's ion acoustic speed $c_s = {((\gamma_ek_BT_0 + \gamma_pk_BT_0/10)/m_p)}^{1/2}$ is $c_s \approx 4.3 \times 10^5$ m/s, where we assumed that $\gamma_e = 5/3$ and $\gamma_p = 3$. 

The simulation box resolves the spatial interval $-44\le y \le 66$ by $5 \times 10^4$ grid cells and the interval $0 \le x \le 0.66$ by 300 grid cells. Periodic boundary conditions are used along $x$ and reflecting ones along $y$. A dense hot pair cloud fills up the interval $y \le 0$. It consists of positrons and electrons with the number density $3n_0$ and the temperature $10^3T_0$. Effects due to the density ratio have been explored in Ref. \cite{Dieckmann18}. The momentum distribution of all plasma species is initialized with the Maxwellian distribution $f(\mathbf{p})\propto \exp{(-\mathbf{p}^2/(2mk_BT))}$ ($m, k_B:$ particle mass and Boltzmann constant), which is not an equilibrium distribution for the temperature $T=10^3T_0$. Its deviation from the correct J\"uttner distribution is small and the initial momentum distribution will rapidly become non-thermal in the region of interest.

The ambient electrons are resolved by $1.5 \times 10^8$ CPs, which are distributed uniformly across the entire simulation box. The electrons and the positrons of the pair cloud are represented by $1.8 \times 10^8$ CPs each. Both species are distributed uniformly across the interval $-44 \le y \le 0$. 
The protons outside the interval $-4.4 \le y \le 22$ are represented by $2.28 \times 10^8$ CPs. The protons within that interval are resolved by $7.62 \times 10^8$ CPs. We refer to the latter as the well-resolved protons. The simulation time $t_{sim}=90$ is subdivided into $2.1 \times 10^5$ time steps.

\section{Simulation results}

\begin{figure}
\includegraphics[width=\columnwidth]{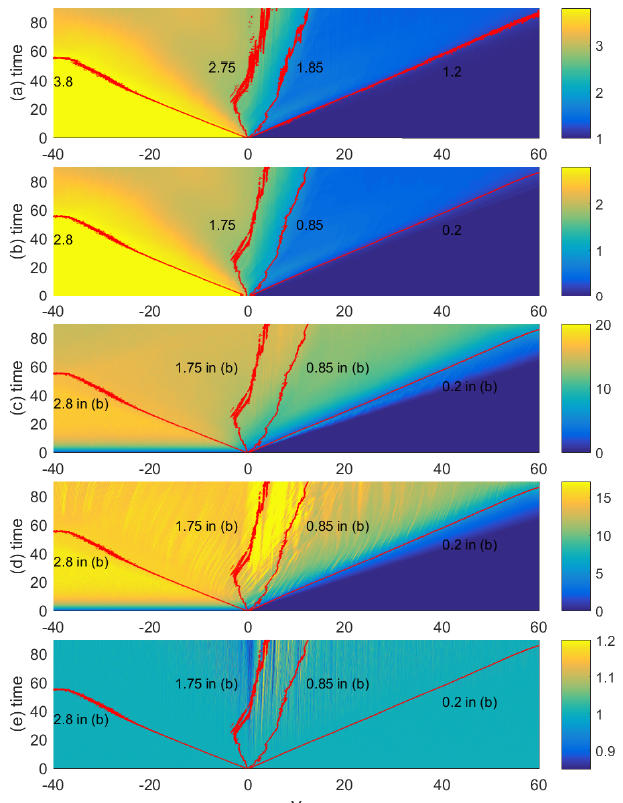}
\caption{The x-averaged densities of the plasma species and of the field energies: panel (a) shows the density distribution of the electrons and (b) that of the positrons. Panel (c) shows the energy density of the in-plane electric field $\epsilon_0(E_x^2+E_y^2)/2$ and (d) that of the out-of-plane magnetic field $B_z^2/2\mu_0$. The proton distribution is shown in (e). Particle densities are normalized to $n_0$ and field energy densities to the thermal pressure $n_0k_BT_0$ of the ambient electrons. The contour lines in (a) are 1.2, 1.85, 2.75, 3.8 and those in all other plots follow the positron density contours 0.2, 0.85, 1.75 and 2.8.}
\label{figure1}
\end{figure}

Figure \ref{figure1} provides an overview of plasma evolution by showing the time-evolution of particle- and field energy densities that have been averaged over $x$. 
The averaged density distributions of the electrons and positrons are qualitatively the same. The electron density is larger by 1 due to the contribution by the ambient electrons. We observe in both distributions a dilute fast moving density structure. Its front is characterized by a density increase of 0.2 and it crosses the distance 60 during the time 80, which corresponds to the speed $0.75c$. The field energy densities in Fig. \ref{figure1}(c,d) increase to about 5-10 times the thermal energy density of the ambient electrons across this beam front. Initially the contours with the density 3.8 in Fig. \ref{figure1}(a) and 2.8 in Fig. \ref{figure1}(b) move to decreasing $y$ at the constant speed 0.8c, reaching $y=-20$ at $t=25$. These contours track the rarefaction wave, which propagates into the pair cloud. 

The contour lines 0.2 and 2.8 in Fig. \ref{figure1}(b) change their slope at late times and close to the box boundaries, which is an artifact introduced by the reflecting boundaries. The plasma evolution close to $y=0$ is, however, not influenced by finite box effects because even the shorter distance from $y=0$ to $y=-44$ and back can only be crossed by the ordinary electromagnetic mode during $t=90$ but not by the slower charge density waves that strongly interact with plasma.

The density distribution in Fig. \ref{figure1}(b) has a significant density gradient between $y\approx 0$ (contour value 1.75) and the contour line 0.85. The latter propagates from $y=0$ to $y\approx 10$ during the displayed time interval, which amounts to a speed $\approx c/10$. The same density distribution close to $y=0$ is observed in Fig \ref{figure1}(a). The bulk of the hot pair cloud thus expands at the speed 0.1c, which amounts to $70 c_s$ in the ambient plasma. 

Figure \ref{figure1}(d) reveals the growth of strong magnetic fields after $t=25$ in the spatial interval occupied by the slowly moving cloud front. No significant change of the electric field's energy density can be seen from Fig. \ref{figure1}(c), which suggests that this is a magnetic instability. The magnetic field distribution it drives is quasi-stationary. The magnetic pressure exceeds the initial proton thermal pressure by the factor 200 and it is thus high enough to modulate the proton distribution in Fig. \ref{figure1}(e). 

\begin{figure}
\includegraphics[width=\columnwidth]{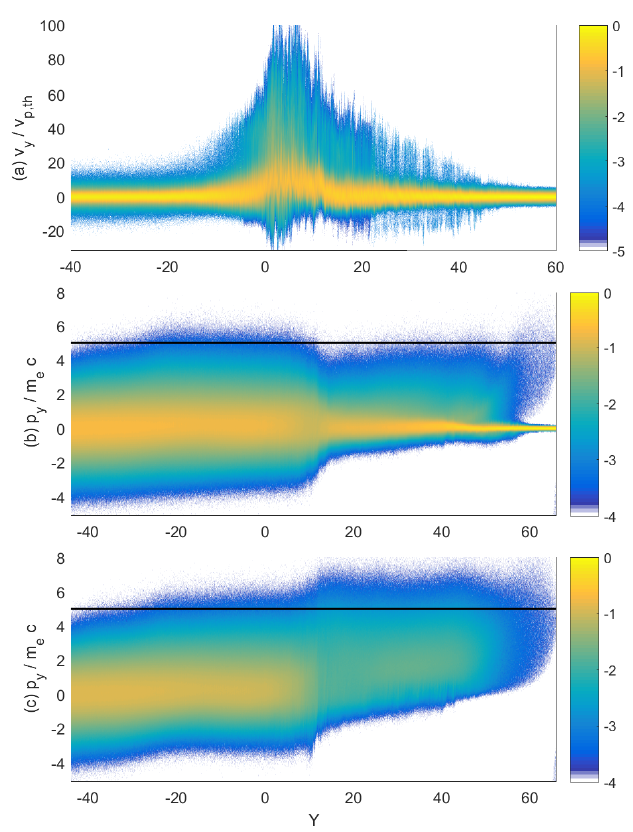}
\caption{The phase space density and momentum distributions of the particles at $t=68$: panel (a) shows the proton distribution normalized to its peak value. The velocity $v_y$ is given in units of the initial proton thermal speed $v_{p,th}$. Panels (b) and (c) show the distributions of the electrons and positrons, respectively. Both are normalized to the peak value in (b). The horizontal lines denote $p_y / m_e c = 5$ and they visualize the difference between the electron- and positron distributions. All densities are displayed on a 10-logarithmic scale (Multimedia view).}
\label{figure2}
\end{figure}

Figure \ref{figure2} shows the x-averaged phase space density distributions of the particles at the time $t=68$.
Figure \ref{figure2}(a) evidences a strong reaction of the protons to the pair cloud's expansion. The bulk population of the protons has been heated and accelerated to increasing $y$ in the interval $0 \le y \le 12$. The apparent mean velocity of the proton's bulk distribution is about 5-10 times $v_{p,th}$ in this interval. The front of the heated protons coincides with the position where the density of the pair cloud decreases.

The phase space density distributions of the electrons and positrons shown by Figs. \ref{figure2}(b,c) change qualitatively across this proton structure. Their momentum distributions, which are no longer symmetric Maxwellians centered at $p_y=0$, follow each other closely for $y\le 0$. Both distributions have been depleted at $p_y/m_ec \le -4$ and $y<0$. The electrons and positrons with a large negative speed moved rapidly to lower $y$, they were reflected by the wall at $y\approx -44$ and now they form the energetic pair population at $p_y/m_ec \ge 4$. The positrons in the interval $y>10$ are hotter than the electrons and they gained momentum along $y$ as they crossed the cloud front. The fastest electrons and positrons have just reached the boundary at $y=66$ at $t=68$ and the $v_y$ component of these particles was thus on average $v_y = 0.97c$. The particle number density increases with decreasing $y$ and reaches its maximum value at $y\approx 50$. The front of the beam's dense part along $y$ moves thus at the speed 0.75 c. This front is responsible for the fast-moving density wave in Figs. \ref{figure1}(a,b). Both beams together form a pair beam with a front that moves with the relativistic factor $\Gamma \approx 1.5-4$. 

\begin{figure}
\includegraphics[width=\columnwidth]{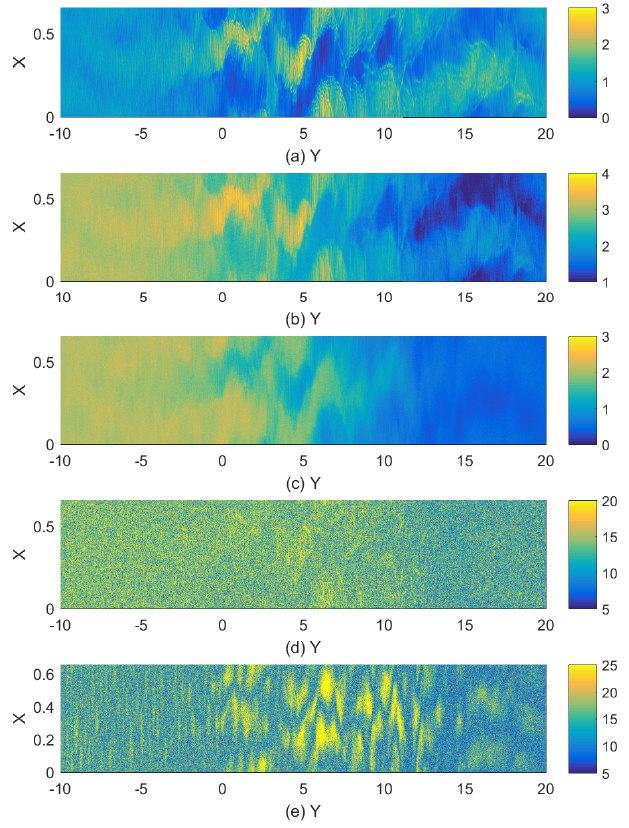}
\caption{The spatial density distributions at the time $t=68$: panel (a) shows the proton density. Panel (b) and (c) show the electron and positron distributions. The energy density of the in-plane electric field is shown in (d) and that of the out-of-plane magnetic field in (e). The field energy densities are normalized to $n_0k_BT_0$ (Multimedia view).}
\label{figure3}
\end{figure}

Figure \ref{figure3} shows the spatial distributions of the particle densities and of the field energy densities at the time $t=68$ close to the center of the simulation box.
The proton density displayed in Fig. \ref{figure3}(a) reveals the presence of filaments, which are approximately aligned with the y-direction. The density oscillations along $x$ reach amplitudes, which are comparable to unity. The proton density oscillations are closely followed by oscillations in the electron density shown in Fig. \ref{figure3}(b). Filaments are also visible in the positron density distribution in Fig. \ref{figure3}(c). Their oscillation amplitude is below that of the protons and electrons and the filaments are shifted relative to those in the electron and proton distributions. The field energy density of the electric field is weak and dominated by random noise at this time. Figure \ref{figure3}(d) (Multimedia view) shows that a two-stream instability developed at early times between the pair cloud and the ambient electrons but that the electric fields it drove were weak at $t=68$. Figure \ref{figure3}(e) shows patches with an energy density of the magnetic field that exceeds by far the thermal one of the ambient electrons. The peak energy density remains, however, below one per cent of that of the pair cloud. Figure \ref{figure3} (Multimedia view) suggests that the instability, which drives the magnetic field and results in the particle filaments, involves all particle species. The magnetic field growth is caused by the spatial separation of electrons and positrons, which move at a mildly relativistic speed to increasing $y$. 

The presence of one density oscillation in Figs. \ref{figure3}(a-c) indicates that the simulation box size along x is just large enough to resolve the wave with the largest unstable wave number. The observed instability thus drives filaments with a scale size comparable to or larger than an proton skin depth. In contrast, Weibel-type instabilities \cite{Weibel59} or filamentation instabilities \cite{Bret10} that involve only electrons or positrons grow during electron time scales and on spatial scales comparable to an electron skin depth ${(m_e/m_p)}^{1/2}$. The involvement of the protons would explain why the instability grows over tens of inverse proton plasma frequencies and why the filaments are so large.

The Weibel instability does not grow if the particle distributions are isotropic in velocity space and filamentation instabilities only grow if the relative speed between beams is comparable or larger than the thermal velocity spread of the beams. We have integrated the distributions over all $x$ and over the interval $-4.4 \le y \le 22$ in order to determine if the electrons and positrons could give rise to an instability on their own.

\begin{figure*}
\includegraphics[width=\textwidth]{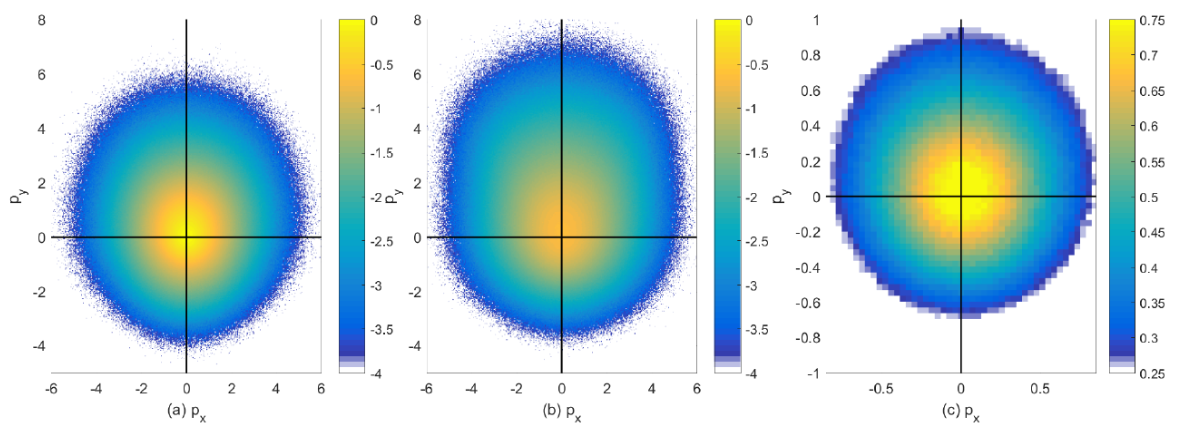}
\caption{The electron and positron momentum distributions averaged over $x$ and over $-4.4\le y \le 22$ at the time $t=68$: panel (a) shows the electron distribution $f_e(p_x,p_y)$ and panel (b) the positron distribution $f_p(p_x,p_y)$ on a 10-logarithmic scale. Both distributions are normalized to the peak value in (a). Panel (c) shows the distribution $f_e(p_x,p_y)-f_p(p_x,p_y)$ on a linear scale (Multimedia view).}\label{figure4}
\end{figure*}
The distributions of the electrons and positrons in Figs. \ref{figure4}(a,b) are hot and almost isotropic. Although some particles have relativistic speeds, the dense core population is nonrelativistic. A beam-Weibel instability can not develop between the electrons and positrons because the thermal spread of the pair cloud is large compared to their relative speed and because electrons and positrons drift in the same direction. The momentum distribution of electrons and positrons is also practically isotropic in the shown plane and such a distribution is unlikely to result in a Weibel instability. Figure \ref{figure4}(b) evidences however a mildly relativistic net drift between the positrons and the protons, which form a point distribution at $p_x, p_y =0$ on this scale, that could give rise to a magneto-instability between both species.

\begin{figure}
\includegraphics[width=\columnwidth]{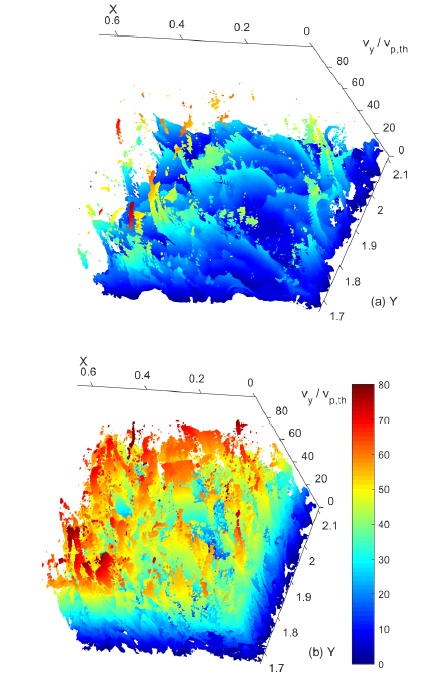}

\caption{Isosurfaces 0.025 (a) and 0.005 (b) of the proton density, which is normalized to its peak density at the time $t=65$. The color denotes velocity $v_y$ in units of $v_{p,th}$ (Multimedia view).}
\label{figure5}
\end{figure}

In what follows we examine the proton distribution in order to better understand how the protons in Fig. \ref{figure2}(a) are accelerated. Figure \ref{figure5} shows two isosurfaces of the proton density distribution $1.7 \le y \le 2.1$.
The isosurface depicting the higher density reveals waves with the amplitude $\sim 20 v_{p,th}$ and wave length $\sim 0.2$. These waves are the equivalent of ion acoustic waves in a pair-ion-plasma. We observe spatially localized structures of accelerating protons. The spatial scale of these acceleration sites is an electron skin depth $\approx 0.023$. Figure \ref{figure5}(b) reveals small dilute clouds of energetic protons. Many are co-located with the accelerating protons in Fig. \ref{figure5}(a) and are thus accelerated by the same process. Once they cease to be accelerated they maintain their large kinetic energy and diffuse out, forming the extended proton clouds at high speeds in Fig. \ref{figure5}(b). Figure \ref{figure5} (Multimedia view) tracks the accelerating protons over the entire range in $y$ in which they are accelerated.

\begin{figure}
\includegraphics[width=\columnwidth]{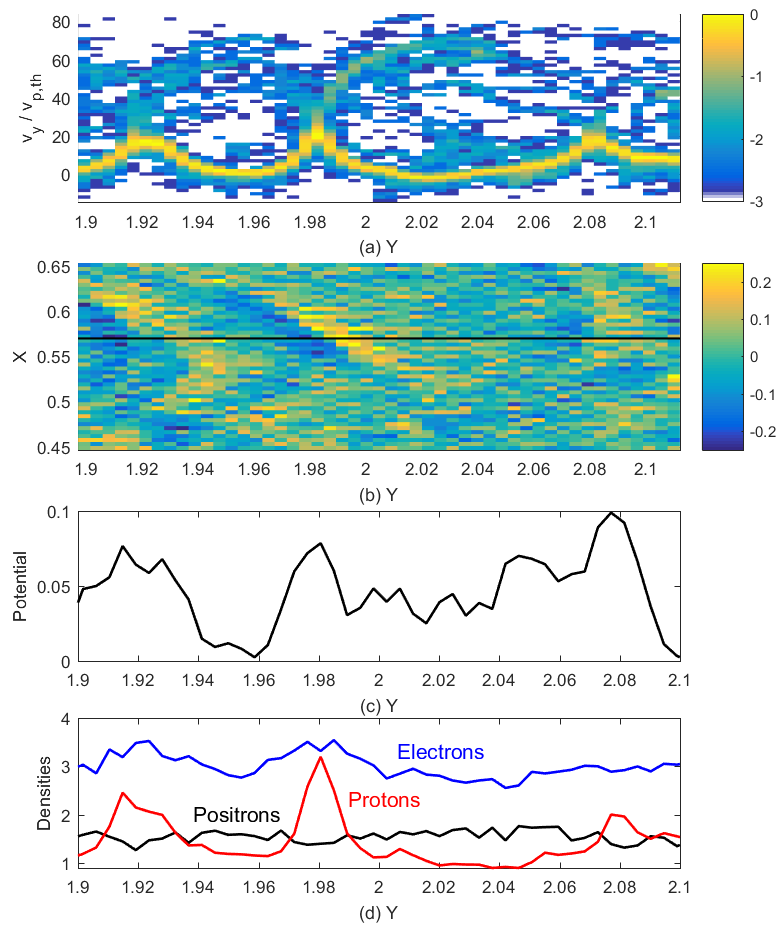}
\caption{Panel (a) shows the proton phase space density along $x=0.57$, which is averaged over 2 cells along $x$ and along $y$, and displayed on a 10-logarithmic scale. Panel (b) shows the electric field $E_y$ in units of $m_e \omega_{pe}c / e$ in a sub-interval around $x=0.57$ (black line). The electric field is averaged over 2 cells along x and along y. Panel (c) shows a lineout of the electric potential along the black line in (b) in units of $m_ec^2$. Panel (d) shows the densities of electrons, positrons and protons integrated over an interval of width $\delta x = 0.026$ centered at $x=0.57$. The time is $t=65$.}
\label{figure6}
\end{figure}

Figure \ref{figure6} displays the electric field $E_y$ in a small interval at $t=65$. The proton phase space density distribution, the electric potential and the densities of the three species along the slice $x=0.57$ are also shown.
Figure \ref{figure6}(a) shows large velocity oscillations of the dense proton core population. The oscillations are not sinusoidal and not centered at $v_y = 0$. The positive mean velocity along $y$ of the oscillating proton core population is responsible for the apparent bulk motion of the protons in Fig. \ref{figure2}(a). Protons are accelerated at the crests of the oscillations. The proton phase space structure resembles that of a breaking ion acoustic wave \cite{Alikhanov70}. Figure \ref{figure6}(b) shows strong electric fields at the locations where protons are accelerated. The electric field distribution shows a bipolar structure at $x=0.57$ and $y\approx 1.98$. 

The amplitude of the oscillations of the potential $U(y)=\int_{y=1.9}^{y} -E_y dy$ in Fig. \ref{figure6}(c) is large enough to change the kinetic energy of an electron by a few percent of $m_e c^2$ and this potential moves with the speed $\approx 15v_{p,th}$ or $0.05c$ to increasing $y$ (See speed of the cusp at $y\approx 1.98$ in Fig. \ref{figure6}(a)). Such potential oscillations hardly affect the relativistically moving electrons of the pair cloud. However, they can trap the electrons of the hot cloud that move at the lowest speed in the box frame and also the ambient electrons as demonstrated in a one-dimensional simulation that resolved only y \cite{Dieckmann18}. The density distributions of the plasma species in Fig. \ref{figure6}(d) shows that the potential oscillation at $y=1.98$ is tied to a large proton density peak. It coincides with a small depletion of the positrons and a maximum of the electron density. The large electron temperature implies that they can not closely follow the proton density and hence the electron density maximum is broad.

The electric field distribution in Fig. \ref{figure6}(b) is typical for an ion acoustic solitary wave like the one in Fig. \ref{figure6}(a) and also for an electron phase space hole, which corresponds to a localized positive excess charge around which the electrons oscillate. Both non-linear plasma structures often coexist and the ion acoustic wave lends the electron phase space hole additional stability \cite{Jenab17}. Protons at the wave crest at $y\approx 1.98$ in Fig. \ref{figure6}(a) could thus be accelerated by such a hybrid structure. 

\begin{figure*}
\includegraphics[width=\textwidth]{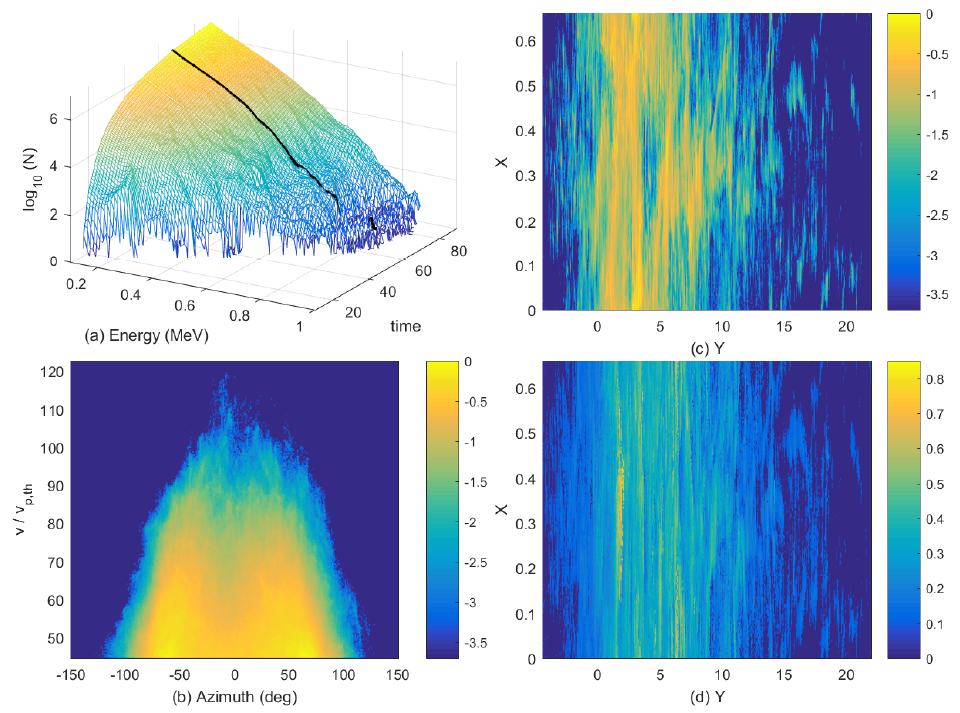}
\caption{The distributions of the well-resolved protons at $t=68$: panel (a) shows the time evolution of the energy spectrum in the interval 0.1 MeV to 1 MeV, which was binned in intervals of width 10 keV. The number of binned protons is $N$ and the black curve follows the distribution at the time $t=68$. Panel (b) shows the 10-logarithmic proton velocity distribution in the simulation plane in polar coordinates and normalized to the peak value at $t=68$. The azimuth angle $\rho$ is defined relative to the y-axis. Panel (c) shows the 10-logarithmic number of energetic protons in units of the initial number of CPs per bin. Panel (d) shows the energy of the most energetic proton in each bin expressed in units of one MeV (Multimedia view).}
\label{figure7}
\end{figure*}

We turn to the analysis of the protons from the well-resolved population with a velocity modulus ${(v_x^2+v_y^2)}^{1/2}>44v_{p,th}$, on a global scale. Figure \ref{figure7} shows various aspects of their energy distribution.
The protons reach kinetic energies above 100 keV after $t\approx 10$ in Fig. \ref{figure7}(a), which is the time when the first proton density modulations appear in Fig. \ref{figure1}(e) but before strong magnetic fields grow in Fig. \ref{figure1}(d). The magnetic field is at least initially not involved in the proton acceleration. The number of high-energy protons rapidly increases after this time and the energy spectrum expands to higher energies. The proton acceleration slows down after $t\approx 60$ and the maximum energy reached by the protons converges to about 850 keV. The slope of the energy spectrum decreases approximately exponentially with the energy. The high-energy component thus follows a Maxwellian distribution over the displayed energy interval. A possible cause for this distribution is the scattering of the protons by many electrostatic structures.

Figure \ref{figure7}(b)  shows that protons increase their energy primarily in the direction of increasing $y$; an azimuth angle $\rho = 0^\circ$ corresponds to the direction of the positive y-axis. The preferential proton acceleration along $y$ matches that observed in Ref. \cite{Dieckmann18}. The cause of this preferential acceleration was that most electron phase space holes propagated in the expansion direction of the pair cloud. The fastest protons with a speed above $10^7$ m/s are found for an azimuth angle of $\rho \approx 0^\circ$. The peak speed decreases with an increasing modulus of the azimuth angle and no energetic protons are observed that propagate at an angle larger than $|\rho|=110^\circ$. Two density maxima are located at $\rho \pm 70^\circ$ and at $v/v_{p,th}\approx 45$ and the density value at these values of $\rho$ remains higher than that of other values of $\rho$ for any given speed up to $80v_{p,th}$. Protons gain energy more easily in the oblique direction while the strongest acceleration takes place along y. Figure \ref{figure7} (multimedia view) shows that the density maxima at oblique angles form at late times. A potential reason for the preferential acceleration in the oblique direction might be that the wave vectors of most electric field structures are not aligned with $y$ (See Fig. \ref{figure6}(b)).

Figure \ref{figure7}(c,d) show the spatial distribution of the energetic protons and of their peak energy. The energetic protons are all located in the interval $0 \le y \le 10$, which coincides with the location where the low-energy protons in Fig. \ref{figure7} reach the highest speed and with the interval in which the pair density in Fig. \ref{figure1}(a,b) decreased. Neither of these plots shows an obvious correlation of the displayed quantity with the proton filaments in Fig. \ref{figure3}(a) or with the magnetic field patches in Fig. \ref{figure3}(e). We do indeed not expect that Larmor rotation of protons plays an important role for their acceleration. The time it takes a proton to complete one rotation is comparable to the simulation time even if we take the maximum observed magnetic field amplitude and assume that it is stationary in space and time. 

\section{Summary}

We have examined the expansion of a hot cloud of electrons and positrons into a diluted electron-proton ambient plasma. The expansion of the pair cloud gave rise to two instabilities. One resulted in the growth of electrostatic ion acoustic solitons well behind the front of the pair cloud and one in a magneto-instability that magnetized the front of the dense part of the pair cloud.

Fluctuations of the number density of the pair cloud and ambient plasma break their initial charge- and current neutrality and instabilities can grow \cite{Stringer64,Gary87,Luque05}. The one-dimensional PIC simulation in Ref. \cite{Dieckmann18} showed that for plasma conditions similar to the ones we use here electron phase space holes form, which are connected to ion solitary waves \cite{Jenab17}. Here we have shown that such solitary waves grow to a large amplitude also in a 2D simulation. The oscillations of their electrostatic potential were large enough to trap the electrons of the background plasma and the low-energy electrons of the pair cloud. The moving potential could also reflect some of the ambient protons \cite{Liseykina15}. The peak energy the reflected protons reached in the two-dimensional simulation was about one third of that in Ref. \cite{Dieckmann18}. 

The bulk of the cloud particles expanded to increasing $y$ at the speed $\sim c/10$. A minor fraction of the pair particles propagated to increasing $y$ at the speed 0.75c. A speed 0.75c is comparable to that of front of the cloud front in Ref. \cite{Dieckmann18}, which did not show a subdivision between a dense slow part and a fast dilute one. The positrons of the fast-moving part of the pair cloud were more energetic than the electrons in both simulations. 

We interpret the subdivision of the pair cloud in the two-dimensional simulation as follows. The cloud particles in the 2D simulation were scattered in the x-y plane by non-planar electromagnetic field structures, which moved at a nonrelativistic speed through the ambient plasma. A one-dimensional geometry prevents the growth of magnetic fields and enforces a planarity of the electric field structures. The particles of the pair cloud undergo a random walk in the x-y plane of the two-dimensional simulation but not in the 1D simulation. 

We observed a magneto-instability at the front of the slow-moving dense part of the pair cloud. The electrons of the pair cloud and of the ambient plasma interacted via a two-stream instability forming a hot electron bath. The hot positrons in the fast outflow had a relativistic mean speed, which could allow them to interact with the ambient protons that were at rest. This instability is similar to the filamentation instability between two ion beams that counterstream at a relativistic speed in a hot electron bath \cite{Yalinewich10}. We observed the formation of density filaments on proton time scales in all particle species,. The filaments formed by electrons and protons were separated by magnetic fields from those that involved positrons. Our simulation box size orthogonal to the flow direction was just big enough to resolve one period of such an instability; this instability is thus likely to result in filaments with a thickness comparable to or larger than an proton skin depth. Finite box effects helped us to identify that an instability grows; future simulations that resolve a larger interval orthogonal to the beam flow direction can study the interplay of these filaments.

\textbf{Acknowledgements:} the simulations were performed on resources provided by the Swedish National Infrastructure for Computing (SNIC) at HPC2N (Ume\aa) and on resources provided by the Grand Equipement National de Calcul Intensif (GENCI) through grant x2016046960. The EPOCH code has been developed with support from EPSRC (grant No: EP/P02212X/1). G. Sarri wishes to acknowledge support from the Engineering and Physical Sciences Research Council (grant number: EP/N027175/1).

\end{document}